\documentclass{appolb}
\usepackage{graphics,fancyhdr,ifthen,amsmath,amssymb,cite,color,graphicx,subfigure}

\begin{document}
\date{\today}
\pagestyle{plain}
\newcount\eLiNe\eLiNe=\inputlineno\advance\eLiNe by -1
\title{Higher harmonic anisotropic flow measurements of charged particles at $\sqrt{s_{_{NN}}} =
$ 2.76 TeV with the ALICE detector
\thanks{Presented at the conference Strangeness in Quark Matter 2011, Cracow, Poland}%
}

\author{You Zhou (for the ALICE Collaboration)
\address{Nikhef, Science Park 105, 1098 XG Amsterdam, The Netherlands\\ 
Utrecht University, P.O.Box 80000, 3508 TA Utrecht, The Netherlands}}
\maketitle

\begin{abstract}
We report the measurements of elliptic flow $v_{2}$, as well as higher harmonics 
triangular flow $v_{3}$ and quadrangular flow $v_{4}$, in $\sqrt{s_{_{NN}}} =$ 2.76 
TeV Pb--Pb collisions,
 measured with the ALICE detector. We show that the measured elliptic and 
triangular flow can be understood from the initial spatial anisotropy and its event--by--event 
fluctuations.
 The resulting fluctuations of $v_{2}$ and $v_{3}$ 
are also discussed.
\end{abstract}

\section{Introduction}
Anisotropic flow is an good observable to study hot and dense matter created in 
heavy-ion collisions. The second order harmonic anisotropic flow $v_{2}$~\cite{JYO-PRD}, 
was studied from SPS
to LHC energies~\cite{SPS-v2, RHIC-v2, LHC-v2} as summarized in~\cite{ART-arXiv}. Recently it has been argued
that due to initial event--by--event geometry fluctuations the third harmonic $v_{3}$, 
called triangular flow, is finite~\cite{BA-PRC}. 
In these proceedings, we will discuss the anisotropic flow and its fluctuations measured
 for charged particles 
in $\sqrt{s_{_{NN}}} =$ 2.76 TeV Pb--Pb collisions.

\section{Data sample and analysis}
For this analysis in these proceedings the ALICE Inner Tracking System (ITS) and the 
Time Projection Chamber (TPC) were used to reconstruct charged particle tracks 
within $|\eta| \textless$ 0.8 and 0.2 $\textless p_{\rm{t}} \textless$ 5.0 GeV/$c$. 
The VZERO counters and the Silicon Pixel Detector (SPD) were used for the trigger.
Only the events whose primary vertex was found within 7 cm from the centre of the detector 
along the beam direction were selected. The tracks are required to have 
at least 70 reconstructed points in the TPC and a $\langle \chi^{2} \rangle$ per TPC 
cluster $\le$ 4. The collision centrality determination utilized the VZERO detectors.
From the study of the collision centrality determined by different
detectors~\cite{Alb-QM}, $\ie$ ZDC, TPC, SPD and VZERO, the centrality resolution is found to be $\textless$ 0.5 $\%$
rms for the most central collisions, while it increases to 2 $\%$ rms for peripheral collisions.

\section{Results and discussion}

\begin{figure}[thb]
\begin{center}
\includegraphics[width=10cm, height=5cm]{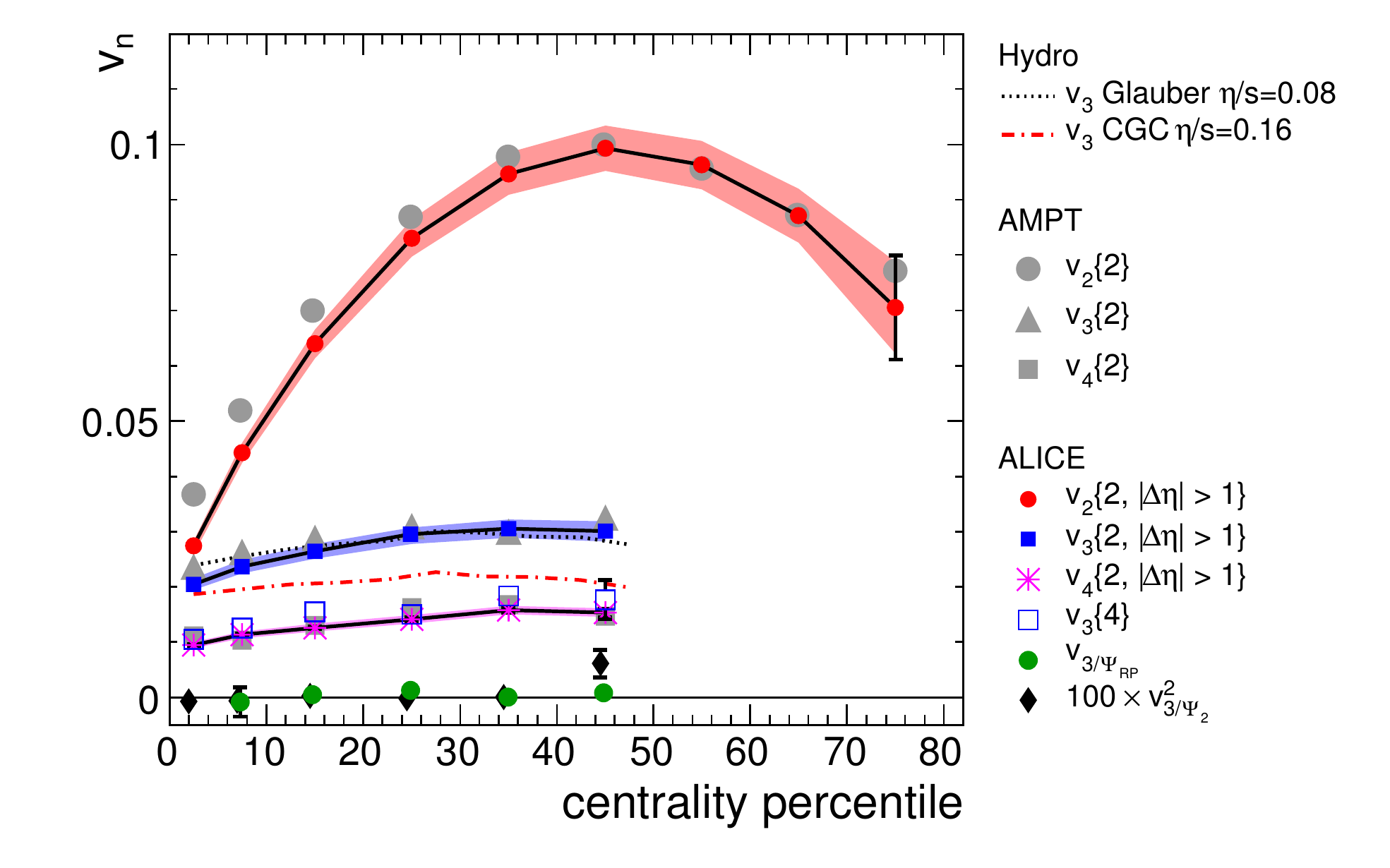}
\caption{$v_{2}$, $v_{3}$ and $v_{4}$ $p_{\rm{t}}$--integrated flow as a function of centrality. 
Full and open blue squares show the $v_{3}\{2\}$ and $v_{3}\{4\}$, respectively. 
The full circle and full diamond are symbols for
$v_{3/\Psi_{\mathrm{RP}}}$ and $v^{2}_{3/\Psi_{2}} $. In addition, the hydrodynamic calculations 
for $v_{3}$ and AMPT simulations are shown by dash lines and full gray markers. 
ALICE data points taken from~\cite{ALICE-V3}.}
\label{Fig1}
\end{center}
\end{figure}

Figure 1 shows the centrality dependence of $v_{2}$, $v_{3}$ and $v_{4}$ integrated over the interval
0.2 $\textless$ $p_{\rm{t}}$ $\textless$ 5.0 GeV/$c$. To suppress non--flow 
effects on the 2--particle cumulant analysis, a minimum $|\Delta\eta|$ gap of one unit 
was used between the correlated 
particles. We correct for the estimated remaining non--flow contributions by using HIJING
~\cite{HIJING}. We observe that the magnitude of $v_{3}$ is much smaller than $v_{2}$ 
(except for the most central collisions)
and does not show a strong centrality dependence. 
These measurements are described by hydrodynamic 
calculations based on Glauber initial conditions and 
$\eta/s = 0.08$, while they are underestimated by hydrodynamic 
calculations with MC--KLN
 initial conditions and $\eta/s = 0.16$~\cite{BHA-PRC}. 
 The comparison suggests a small value 
 of $\eta/s$ for the produced matter. The $v_{3}$ measured from
  the 4--particle cumulant is about
  a factor 2 smaller than the 2--particle cumulant estimate, which can be understood 
  if $v_{3}$ originates predominantly from event--by--event fluctuations 
  of the initial spatial geometry~\cite{RSB-PRC}. At the same time, 
  we evaluate the correlation between 
  $\Psi_{3}$ and the reaction plane 
  $\Psi_{\mathrm{RP}}$ via $v_{3/\Psi_{\mathrm{RP}}} = \langle \cos ( 3 \phi - 
  3 \Psi_{\mathrm{RP}}) \rangle$. In addition
  the correlation of $\Psi_{3}$ and $\Psi_{2}$ also can be studied by a 5--particle correlator 
  $v_{3/\Psi_{2}}^{2} = \langle \cos (3 \phi_{1} + 3 \phi_{2} - 2 \phi_{3} - 2 \phi_{4} - 2 \phi_{5} ) 
  \rangle / v_{2}^{3}$. In Fig. 1 we observe that 
  $v_{3/\Psi_{\mathrm{RP}}}$ and $v_{3/\Psi_{2}}^{2}$
 are consistent with zero within uncertainties. Based on these results, 
 we conclude that $v_{3}$ develops as a correlation of all particles with respect 
 to the third order participant plane $\Psi_{3}$, while there is no (or very weak) 
 correlations between 
 the $\Psi_{\mathrm{RP}}$ (also for $\Psi_{2}$) and the $\Psi_{3}$.
 Finally, from the
  comparison of AMPT model calculations with our measurements, 
  we find that this model can describe the experimental data very well; there is only a slight 
  overestimation of $v_{2}\{2\}$ in the most central collisions~\cite{JX-PRC}.  
 \begin{figure}
\includegraphics[width=6.1cm,height=4.5cm]{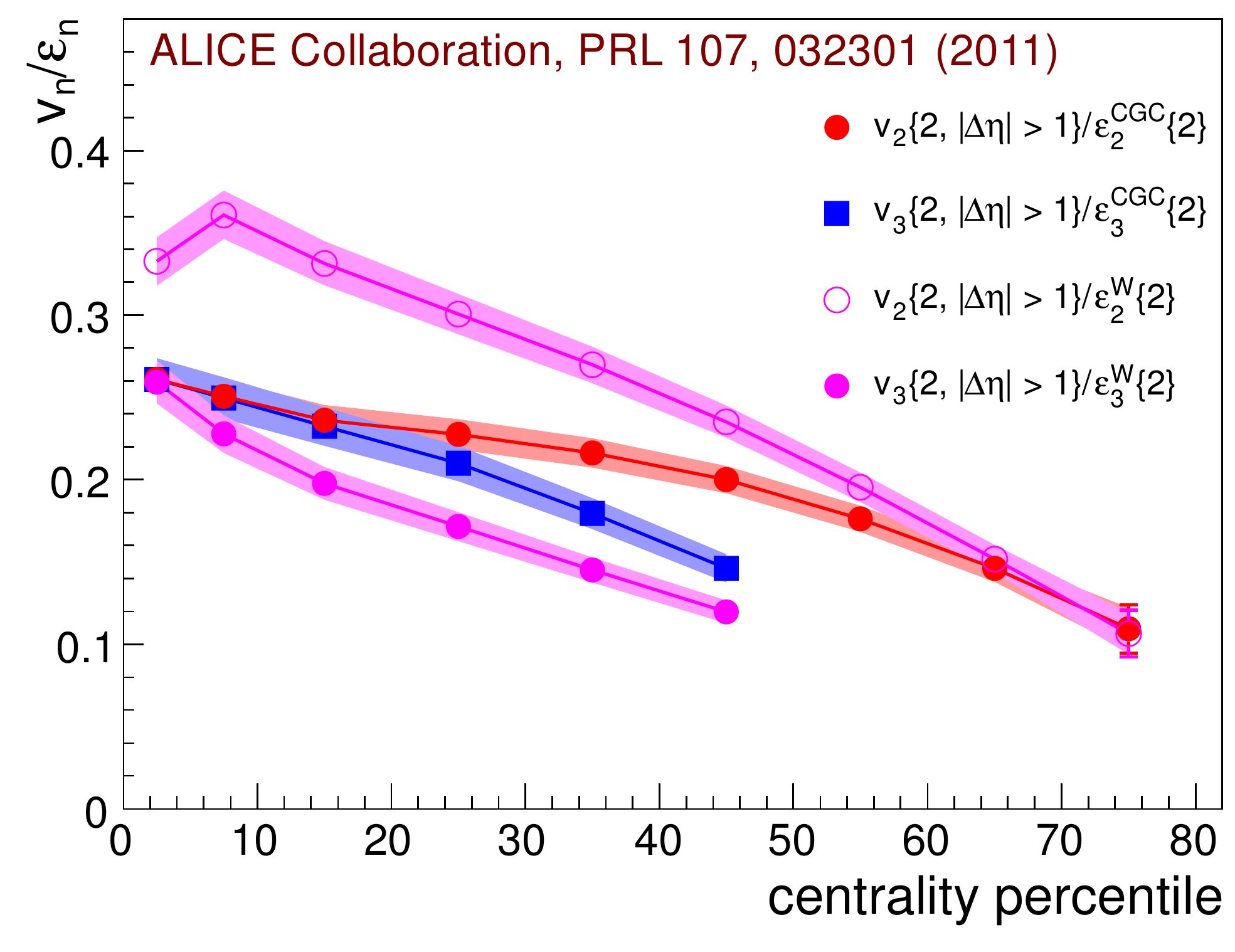}
\includegraphics[width=6.8cm]{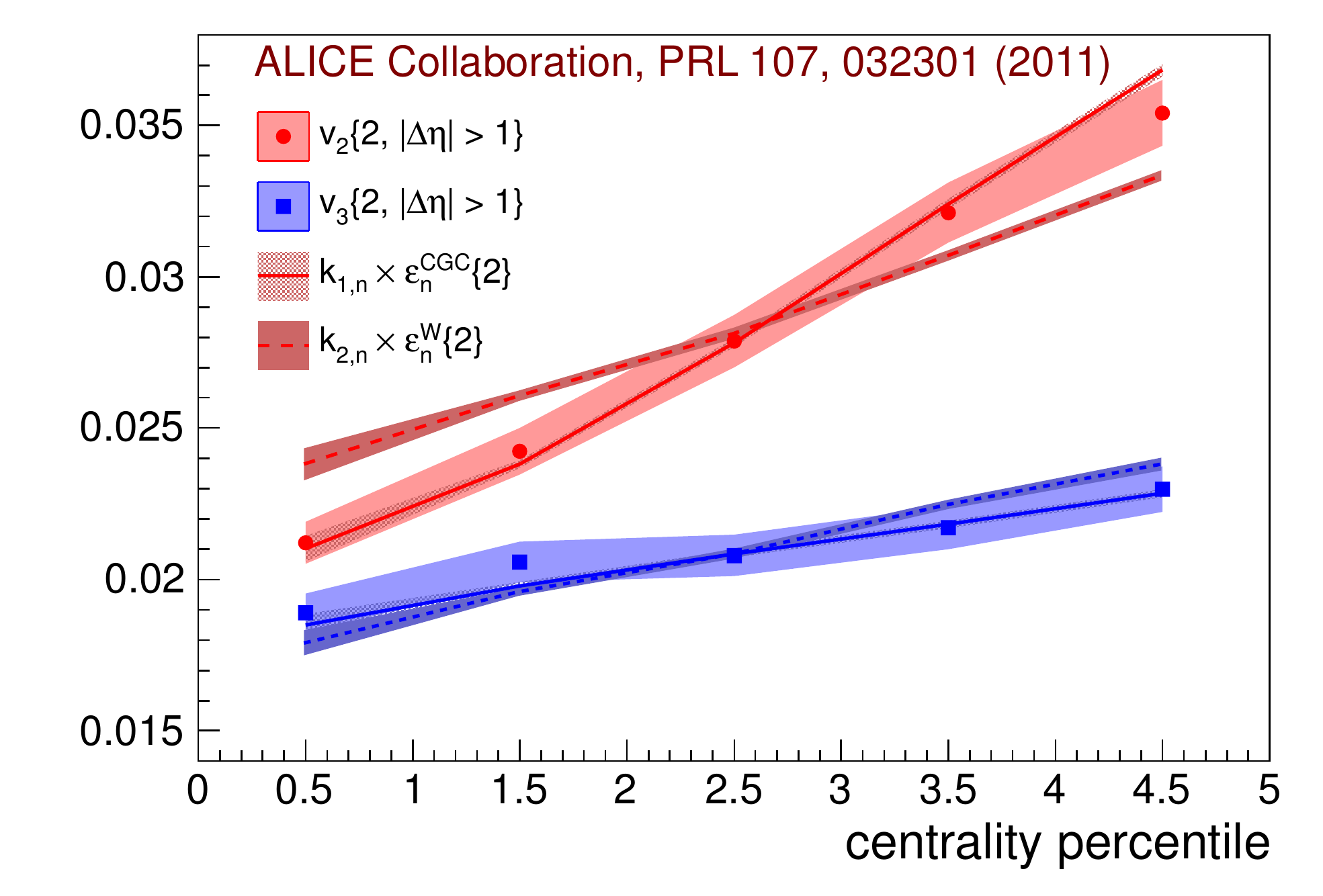}
\caption{(left) The centrality dependence of $v_{2}/\varepsilon_{2}$ and 
$v_{3}/\varepsilon_{3}$ for MC--Glauber and MC--KLN CGC initial conditions.  (right) $v_{n}$ and scaled $\varepsilon_{n}$ as a function of centrality for the most central collisions (0--5$\%$). The $k_{1}$ and
$k_{2}$ have been used for eccentricities to match the 2$\%$ and 3$\%$ centrality percentile data. Figure taken from~\cite{ALICE-V3}.}
\label{Fig2}
\end{figure}

To investigate the role of viscosity on anisotropic flow measurements, 
we calculate the ratio $v_{2}/\varepsilon_{2}$ and 
$v_{3}/\varepsilon_{3}$. Here $\varepsilon_{2}$ and $\varepsilon_{3}$ are the eccentricity 
and triangularity of the initial spatial geometry which are defined by 
\begin{equation}
\varepsilon_{n}= \frac{\sqrt{\langle r^{2} \sin (n \phi) \rangle ^{2} 
+ \langle r^{2} \cos (n \phi) \rangle ^{2} }} {\langle r^{2} \rangle }.
\end{equation}
The definition of $\varepsilon_{n}\{2\}$ (also $\varepsilon_{n}\{4\}$) can be found in~\cite{MM-arXiv}.

Figure 2 (left) shows the centrality dependence of the ratio $v_{n}/\varepsilon_{n}$. The $\varepsilon_{n}$  
are extracted from the MC--Glauber model (using the number of wounded nucleons) 
and the MC--KLN CGC model, denoted by 
 $\varepsilon_{n}^{\mathrm{W}}\{2\}$ and $\varepsilon_{n}^{\mathrm{CGC}}\{2\}$, respectively. 
 Based on the assumption that $v_{n}\propto\varepsilon_{n}$, 
 we get $v_{n}\{2\}\propto\varepsilon_{n}\{2\}$~\cite{MM-arXiv}. 
 We observe that 
 the $v_{2}\{2\}/\varepsilon_{2}^{\mathrm{W}}\{2\}$ is larger than 
 $v_{3}\{2\}/\varepsilon_{3}^{\mathrm{W}}\{2\}$ in all centrality bins,
  which indicates significant viscous corrections. However, for the MC--KLN CGC model 
  the magnitude of $v_{2}\{2\}/\varepsilon_{2}^{\mathrm{CGC}}\{2\}$ equals to
   $v_{3}\{2\}/\varepsilon_{3}^{\mathrm{CGC}}\{2\}$ in the most central collisions, 
   which might be expected for an almost ideal fluid~\cite{BHA-PRC}. 
   The ratio of $v_{3}\{2\}/\varepsilon_{3}^{\mathrm{CGC}}\{2\}$ decreases faster
    than $v_{2}\{2\}/\varepsilon_{2}^{\mathrm{CGC}}\{2\}$ from central to peripheral collisions, 
   which is consistent with larger viscous corrections to $v_{3}$.
 The collective flow should be directly sensitive to the change of the initial spatial geometry 
since the viscous effects do not change too much in the 
small centrality range. In  Fig. 2 (right) we observe 
that in this centrality range $v_{3}$ does not show a strong centrality dependence while the $v_{2}\{2\}$ 
increases significantly. The comparison of the scaled initial eccentricity
shows that $v_{2}\{2\}$ and $v_{3}\{2\}$ can only be simultaneously described by
 $\varepsilon_{2}\{2\}$ and $\varepsilon_{3}\{2\}$ from the MC--KLN model.

 \begin{figure}
\includegraphics[width=6.5cm, height=4.5cm]{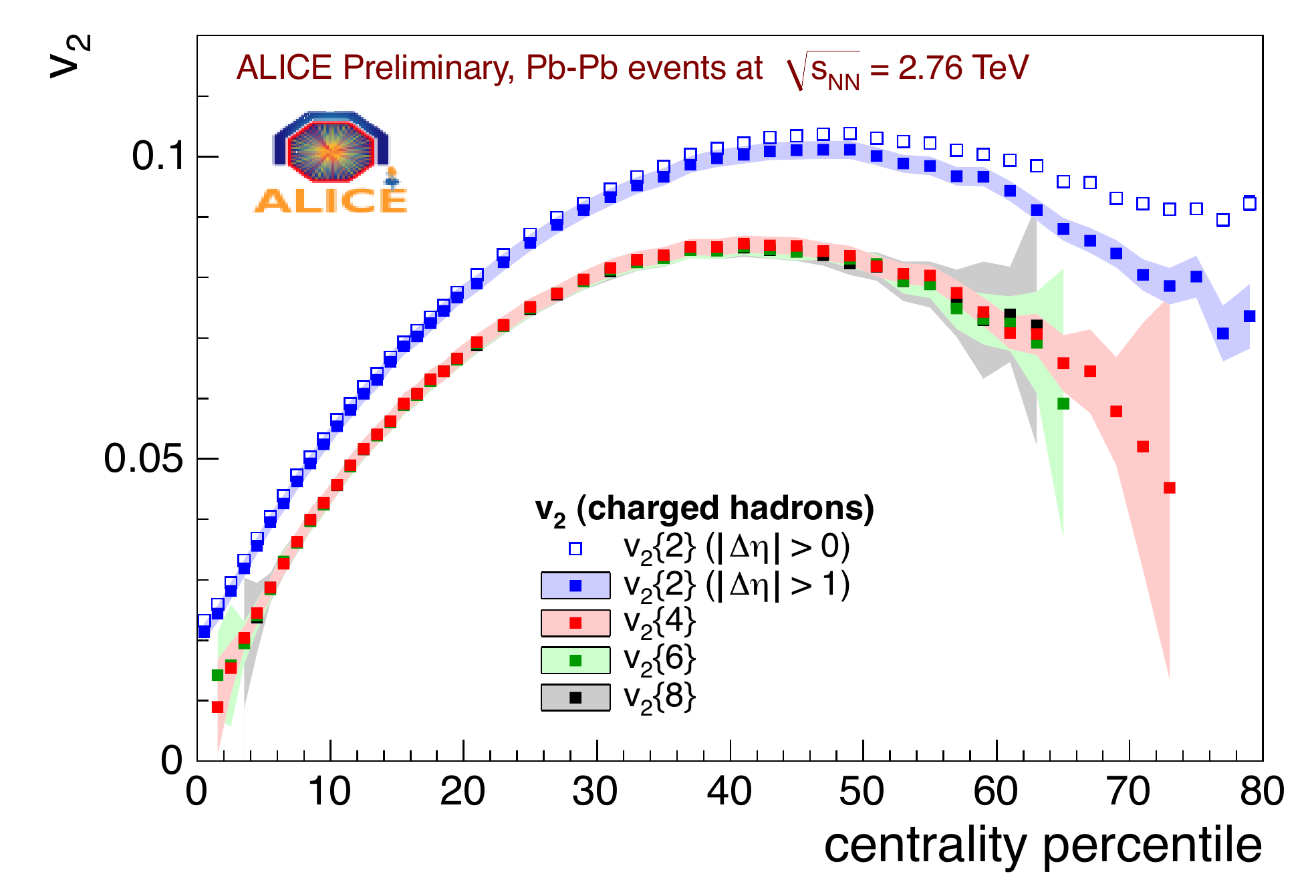}
\includegraphics[width=6.5cm, height=4.5cm]{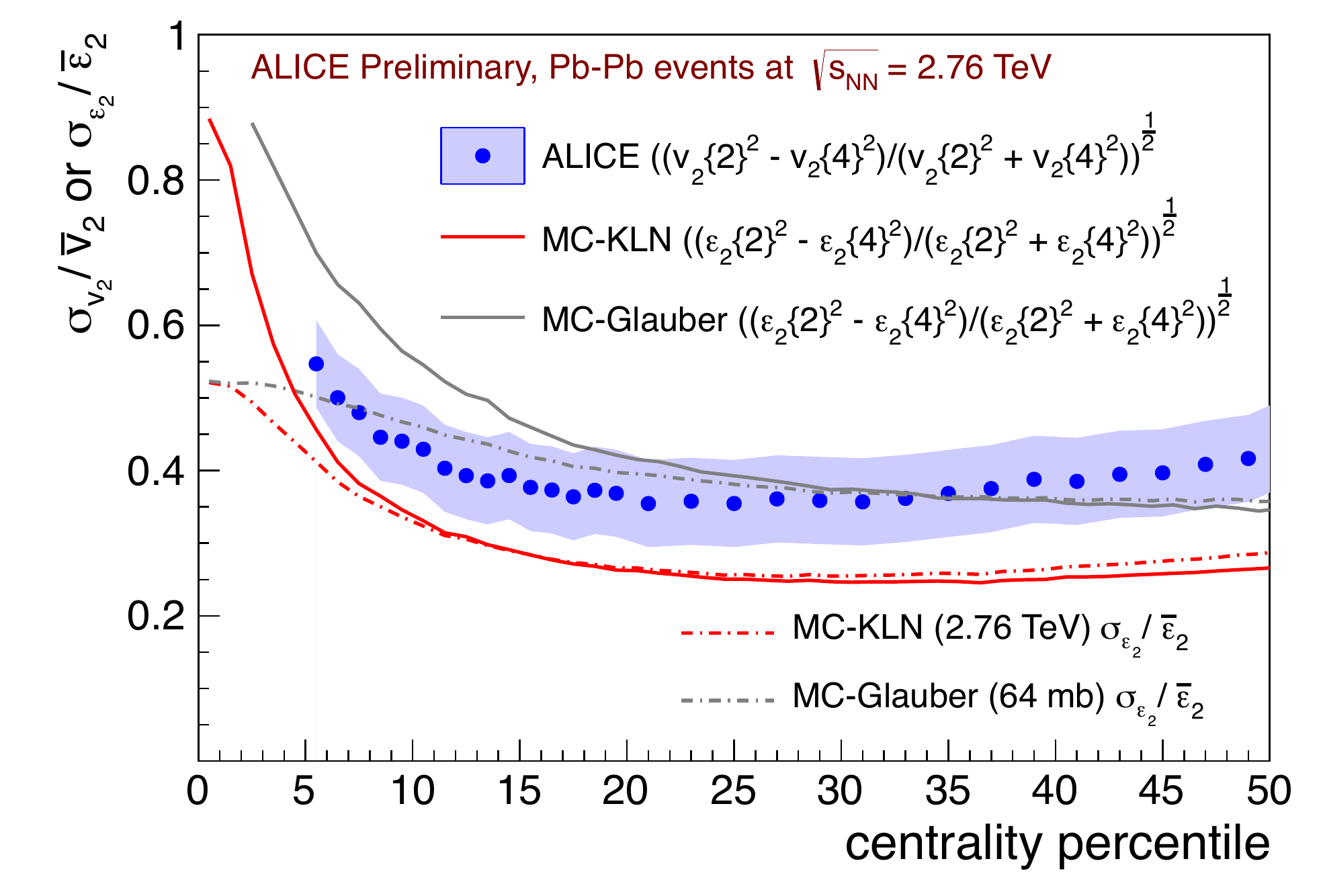}
\caption{(left) The centrality dependence of integrated $v_{2}$ in 1-2 $\%$ bins with different order 
cumulants. (right) Relative flow (eccentricity) fluctuations versus centrality. }
\label{Fig1}
\end{figure}

In order to reduce the event--by--event fluctuations within a centrality bin, 
we plot the  integrated $v_{2}$ as a function of centrality in narrow bins, 1$\%$ centrality bins 
for 0--20$\%$ and 2$\%$ bins for 20--80$\%$~\cite{Alb-QM}. Elliptic flow estimated from 2--particle 
azimuthal correlations, $v_{2}\{2\}$, 
was obtained by using two different pseudorapidity gaps ($|\Delta \eta|\textgreater 0$
 and $|\Delta \eta|\textgreater 1$). 
 The difference between the two measurements can be understood as resulting from non--flow effects. 
 At the same time, the results of 4--, 6-- and 8--particle
  cumulants estimates are shown in Fig. 3 (left). The good agreement of the multi-particle cumulants
  indicates that with 4--particle cumulants non--flow is strongly suppressed, so that there is little gain in 
  suppressing it further by higher order cumulants (like 6-- and 8--particle).

 As shown by Gaussian fluctuations studies~\cite{SV-FF}, in the limit of small fluctuations
   ($\sigma_{v} \textless \bar{v}$), we can estimate the participant plane flow and its fluctuations with:
  \begin{equation}
\bar{v}_{n} \approx \sqrt{\frac{v_{n}^{2}\{2\} + v_{n}^{2}\{4\}}{2}} ~
\mathrm{and} ~\sigma_{v_{2}} \approx \sqrt{\frac{v_{2}^{2}\{2\} - v_{2}^{2}\{4\}}{2}}.
\end{equation}
However, in the case of only fluctuations~\cite{SV-FF}, we have 
 \begin{equation}
 \bar{v}_{n} = \frac{\sqrt{\pi}}{2}v_{n}\{2\}
~(\mathrm{or}~ \sigma_{v_{n}}/\bar{v}_{n}=\sqrt{4/\pi-1}) 
~\mathrm{and} ~v_{n}\{4\} =0.
\end{equation}
Also based on the assumption that $v_{n}$ is proportional to $\varepsilon_{n}$, 
the centrality dependence of eccentricity and its fluctuations should 
show behavior similar to that of flow.
In Fig. 4 (left) we indeed observe a similar centrality dependence of 
$\varepsilon_{2}$ with $v_{2}$ and the following equations are valid
\begin{equation}
\varepsilon_{2}^{2}\{2\} \approx \varepsilon_{2}^{2} + \sigma_{\varepsilon_{2}}^{2}, ~
\varepsilon_{2}^{2}\{4\} \approx \varepsilon_{2}^{2} - \sigma_{\varepsilon_{2}}^{2}
\end{equation}
with the exception of the most central collisions (for which $\sigma_{v} \textless \bar{v}$ does not hold).

\begin{figure}[thb]
\includegraphics[width=6.5cm, height=6cm]{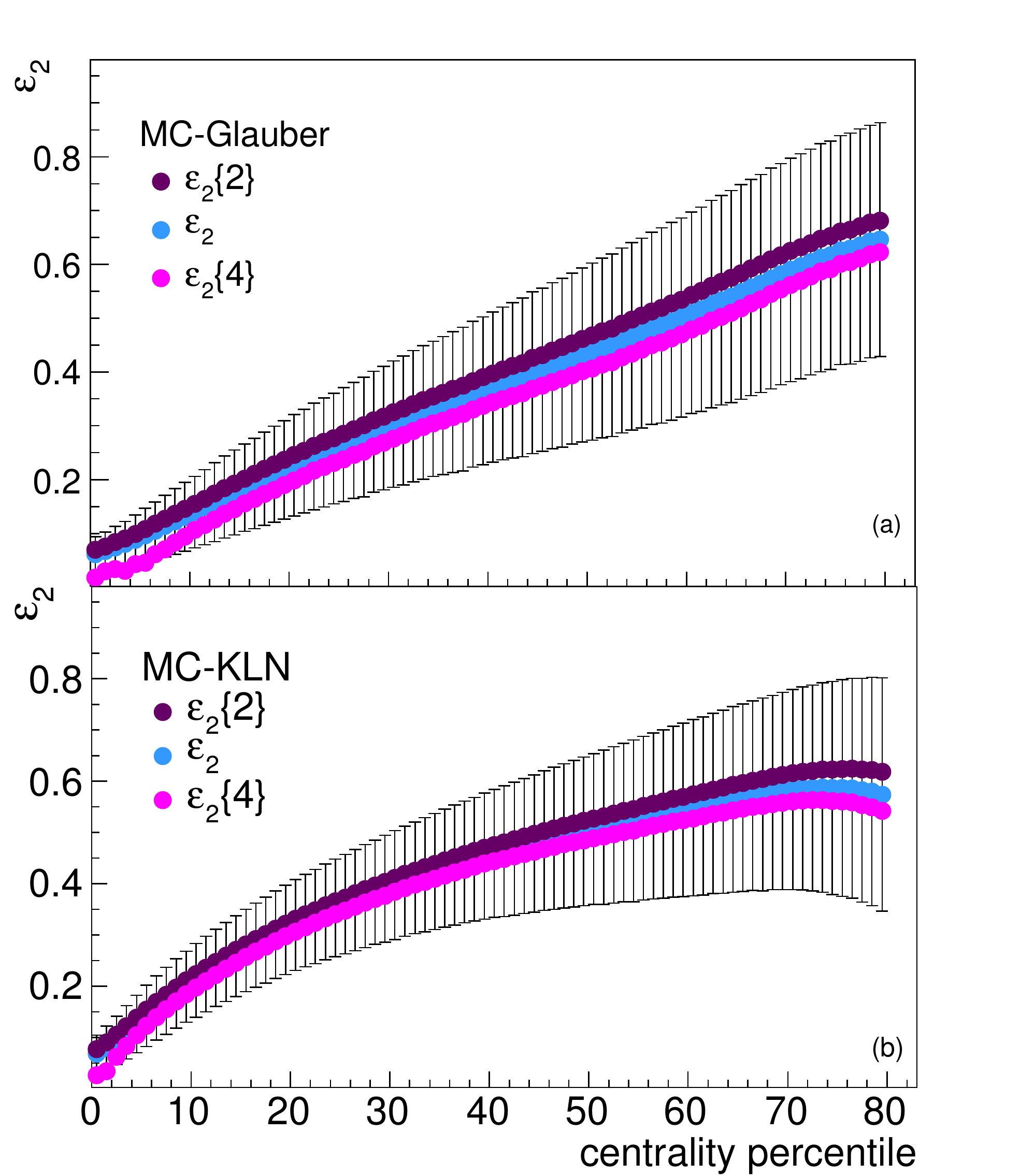}
\includegraphics[width=6.5cm, height=6cm]{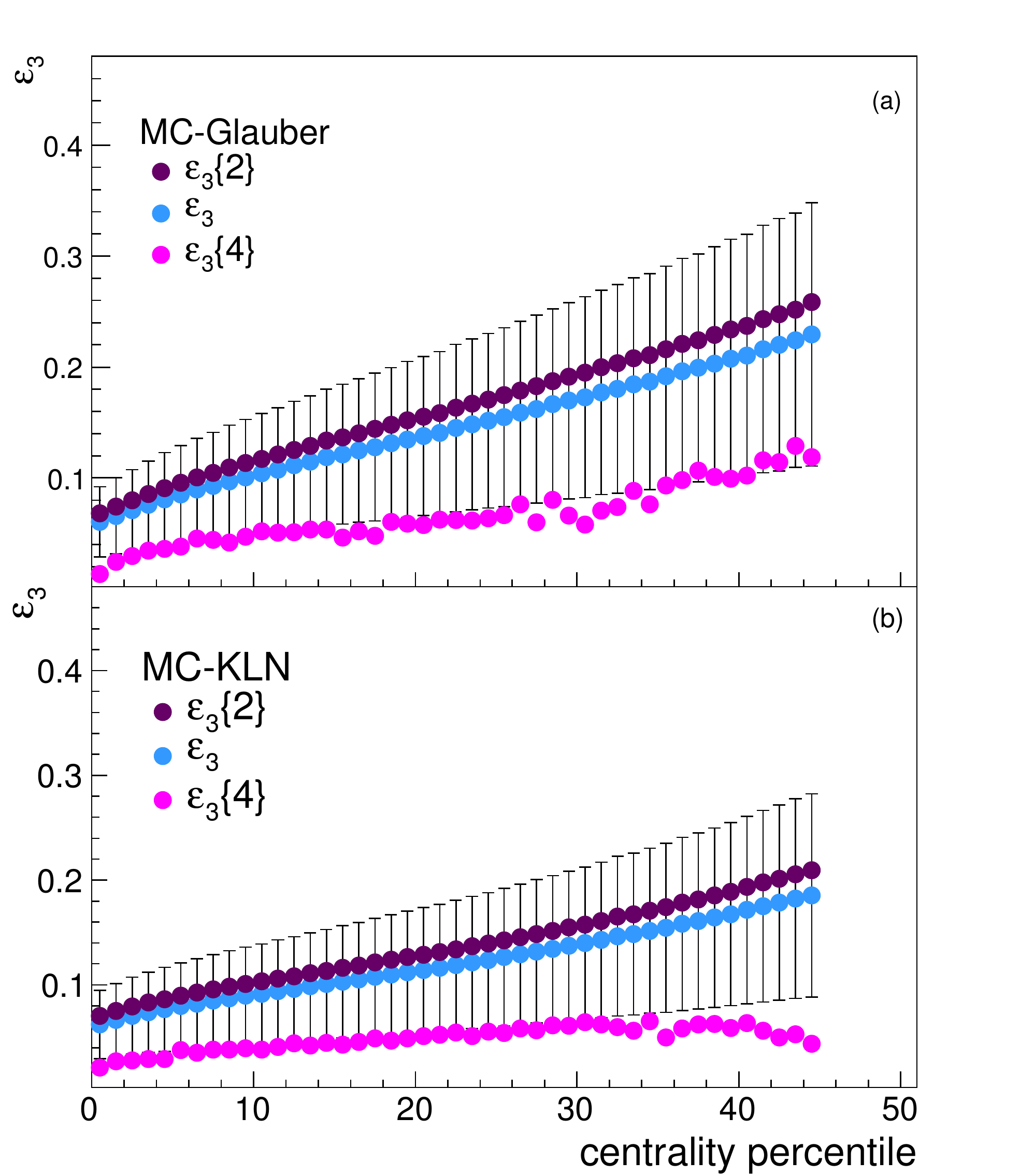}
\caption{Centrality dependence of eccentricity (left) and triangularity (right) from 
MC--Glauber model and MC--KLN model. The error bar is the fluctuations of the 
eccentricity (triangularity). The $\varepsilon_{n}$ is extracted from the Eqs. (1), 
the definitions of $\varepsilon_{n}\{2\}$ and $\varepsilon_{n}\{4\}$ can be found in~\cite{MM-arXiv}.}
\label{Fig1}
\end{figure}
 The centrality dependence of the relative flow fluctuations $\sigma_{v_{2}}/\bar{v}_{2}$
  are plotted in Fig. 3 (right).
 We find that the magnitude of relative flow fluctuations is around 40$\%$.
Also we show the comparison of the relative flow fluctuations to 
$\sigma_{\varepsilon_{2}}/\varepsilon_{2}$ (extracted from Eqs. (4)) 
from both the MC--Glauber model and the MC--KLN model.
 In mid-central and mid-peripheral collisions, the MC--Glauber model can describe
 the flow fluctuations while the MC--KLN model underestimates the measurements. 
 In the more central collisions, neither the MC--Glauber nor the MC--KLN can describe the data.
At the same time, we notice that $\sigma_{\varepsilon_{2}}/\varepsilon_{2}$ from both
the MC--Glauber model and the MC--KLN model reach $\sqrt{4/\pi -1}$ in the most central collisions, 
which is consistent with the predictions if there are only fluctuations~\cite{SV-FF}.

Assuming that $v_{3}$ originates from the initial geometry fluctuations 
(there are only flow fluctuations), we expect that 
$v_{n}\{2\} =\frac{2}{\sqrt{\pi}}\bar{v}_{n}~\mathrm{and} ~v_{n}\{4\} =0$. 
However, as we have shown in Fig. 1, the $v_{3}\{4\}$ has a finite magnitude. 
 In order to understand the fluctuations of $v_{3}$, we look at the centrality dependence of 
triangularity $\varepsilon_{3}$.
 In Fig. 4 we observe that $\varepsilon_{3}\{2\} =\frac{2}{\sqrt{\pi}}\varepsilon_{3}$ is 
 still valid (or
  the ratio $\sigma_{\varepsilon_{3}}/\varepsilon_{3}$ equals to $\sqrt{4/\pi -1}$),
but it seems that $\varepsilon_{3}\{4\} \approx \varepsilon_{3}-\sigma_{\varepsilon_{3}}$ 
in the centrality bins we present here. 
Whether the fluctuations of $\varepsilon_{3}$ are the dominant contribution to  the fluctuations of $v_{3}$
is currently unknown.

\section{Summary}
\parindent=0pt

In these proceedings we have presented the results on anisotropic flow measured in Pb--Pb
collisions at $\sqrt{s_{_{\mathrm{NN}}}}$=2.76 TeV by ALICE at the LHC. The measurements of 
higher harmonic anisotropic flow, in particular $v_{3}$, provide new constraints on the initial
anisotropy as well as the shear viscosity to entropy density ratio $\eta/s$.




\end{document}